\documentclass[twocolumn,preprintnumbers,amsmath,amssymb,pra,showpacs]{revtex4-1}

\usepackage{graphicx}
\usepackage{dcolumn}
\usepackage{bm}
\usepackage{bbm}
\usepackage{amsfonts}
\usepackage{hyperref}
\usepackage{color}
\usepackage{multirow}

\newcommand{\ket}[1]{| #1 \rangle}
\newcommand{\bra}[1]{\langle #1 |}

\newcommand{\rt}[1]{\frac{1}{\sqrt{#1}}}
\newcommand{\fr}[1]{\frac{1}{#1}}
\newcommand{\ketbra}[2]{| #1 \rangle \langle #2 |}

\newcommand{\sing}{\ket{\psi^-}}
\newcommand{\singb}{\bra{\psi^-}}
\newcommand{\cre}[1]{#1^\dagger}
\newcommand{\cru}[1]{\hat{a}^\dagger_{#1}}

\newcommand{\crusq}[1]{\hat{a}^{\dagger 2}_#1}
\newcommand{\da}{^\dagger}
\newcommand{\eye}{\mathbbm{1}}

\begin{document}

\title{Optical spin-1 chain and its use as a quantum computational wire}
\author{Andrew S. Darmawan}%
\author{Stephen D. Bartlett}
\affiliation{School of Physics, The University of Sydney, Sydney,
New South Wales 2006, Australia}%

\date{26 July 2010}


\begin{abstract}
Measurement-based quantum computing, a powerful alternative to the standard circuit model, proceeds using only local adaptive measurements on a highly-entangled resource state of many spins on a graph or lattice.  Along with the canonical cluster state, the valence-bond solid ground state on a chain of spin-1 particles, studied by Affleck, Kennedy, Lieb, and Tasaki (AKLT), is such a resource state.  We propose a simulation of this AKLT state using linear optics, wherein we can make use of the high-fidelity projective measurements that are commonplace in quantum optical experiments, and describe how quantum logic gates can be performed on this chain.  In our proposed implementation, the spin-1 particles comprizing the AKLT state are encoded on polarization biphotons: three level systems consisting of pairs of polarized photons in the same spatio-temporal mode. A logical qubit encoded on the photonic AKLT state can be initialized, read out and have an arbitrary single qubit unitary applied to it by performing projective measurements on the constituent biphotons.  For MBQC, biphoton measurements are required which cannot be deterministically performed using only linear optics and photodetection.
\end{abstract}

\maketitle

\section{Introduction}

In recent years, a significant amount of research has been dedicated to overcoming the practical hurdles posed by the full-scale realization of quantum computers. Measurement-based quantum computation (MBQC)~\cite{raussendorf2001} is an alternative model to the standard circuit model~\cite{nielsen2000} that significantly reduces the requirements for quantum computation in a number of architectures including linear optics~\cite{nielsen2004,browne2005}. In MBQC, the computation proceeds by performing single-particle adaptive measurements on a fixed multi-partite entangled ``resource'' state, which is defined on a set of quantum particles arranged on a graph or lattice. The primary challenge for quantum computation is then shifted from achieving controlled unitary evolution, as in the standard circuit model, to preparing, maintaining and performing measurements on such a resource state.  The \emph{cluster state}~\cite{raussendorf2001} serves as the canonical resource state for MBQC, but recently there have been a few proposed alternatives~\cite{bartlett2006,gross2007a,gross2007,brennen2008,doherty2009,chen2009,barrett2009}.  Among these, perhaps the most intriguing are those that arise as the ground state of a ``natural'' spin-lattice Hamiltonian with two-body nearest-neighbour interactions.  With such a Hamiltonian model, the resource state can be created simply by cooling~\cite{jennings2009}, rather than a complex dynamical construction.

One such proposal~\cite{gross2007,brennen2008} is based on the \emph{AKLT state}, named after Affleck, Kennedy, Lieb and Tasaki~\cite{affleck1987}. The AKLT state is defined on a one-dimensional chain of spin-1 particles and is the ground state of a two-body, rotationally-invariant, nearest-neighbour antiferromagnet
\begin{equation}
H^{\rm AKLT} = \sum_{i=1}^N \vec{S}_i \cdot \vec{S}_{i+1} + \fr{3} (\vec{S}_i \cdot \vec{S}_{i+1})^2\,,
\label{e:akltinfinite}
\end{equation}
where $\vec{S}_i$ is the spin-1 operator acting on the $i$-th site. In condensed matter physics, the AKLT state (an example of a valence-bond solid~\cite{affleck1987}) was put forward as a rigorous example supporting Haldane's conjecture~\cite{haldane1983, haldane1983a} that 1-D Heisenberg chains with integer spins, as opposed to half-integer spins, have a non-zero energy gap.  Along with its role in theoretical condensed matter physics, the AKLT state has served as a template for understanding quantum information processing using spin-chains with a measurement-based model.  The mathematical methods in quantum information theory that were developed from generalizing the AKLT state, such as finitely-correlated states~\cite{fannes1992}, matrix product states~\cite{garcia2007} and projected entangled pair states (PEPS)~\cite{perezgarcia2008}, form the basis for our theoretical description of MBQC and the development of new resource states. The AKLT state with open boundary conditions is a perfect qubit channel, with maximal, infinite-ranged localizable entanglement~\cite{verstraete2004b}.  In fact, it is an even stronger resource than this, as it can serve as a \emph{quantum computational wire}~\cite{gross08}.  A quantum computational wire is a linear multipartite state (e.g., the ground state of a spin chain) that can transmit a logical qubit along its length by performing single particle measurements, and in addition can apply single qubit unitaries to this logical qubit.  Quantum computational wires can be used as basic components of a quantum computer; by coupling multiple such wires together, one can construct universal resources for MBQC~\cite{brennen2008,gross2007}. 

In the condensed-matter systems normally associated with strongly-interacting spin chains, there is currently no way to perform the high-fidelity adaptive measurements of individual spins required for MBQC.  However, considerable recent progress in developing strongly-interacting quantum optical and atomic systems on lattices with controllable interactions may allow us to synthesize such an interaction in an architecture where such measurements are possible.  Brennen and Miyake~\cite{brennen2008} propose possibilities using neutral atoms with controlled collisions in an optical lattice, or polar molecules with dipole-dipole interaction in an optical lattice. 

Here, we propose an experimental method for simulating an AKLT state using single-photon linear optics, with biphotons as the spin-1 particles, and detail its use as a quantum computational wire.  This proposal takes advantage of the high-fidelity projective measurements that are available in single-photon experiments.  Our proposal has many similarities to the linear optical methods used to generate cluster states~\cite{nielsen2004,browne2005}, but also some key differences.  First and foremost, our optical AKLT state uses qutrits (three-level quantum systems) rather than qubits.  Higher dimensional systems, like qutrits, have been shown to possess advantages in quantum information processing, for instance in terms of increased channel capacity \cite{fujiwara2003} and increased security in quantum bit commitment \cite{langford2004}. Biphotons, being qutrits, are natural candidates for these applications, and recent work has illustrated how these biphotons may be manipulated in linear optics. Lanyon \textit{et al.}~\cite{lanyon2008} have experimentally demonstrated how a given input biphoton may be transformed into an arbitrary biphoton, and Lin \cite{lin2009} has shown how arbitrary unitary operations may be applied to biphotons.  We make use of these recent capabilities for linear-optics manipulation of biphotons for our proposal.



The paper is structured as follows.  In Sec.~\ref{s:aklt}, we review the definition of the AKLT state and its description as a matrix product state.  In Sec.~\ref{s:optical}, we present our proposal for an optical implementation of the AKLT state, as well as the methods for using this state as a quantum computational wire.  We conclude with a discussion in Sec.~\ref{s:conclusion}.

\section{The AKLT state}
\label{s:aklt}

We first review some of the basic properties of the AKLT state. Consider a one-dimensional chain of spin-1 particles. The AKLT state is a spin-1 antiferromagnet, and can be defined by requiring that that the total spin of every neighbouring pair of particles is never $J=2$. For an infinite chain the Hamiltonian for which the AKLT state is the ground state may be constructed simply as
\begin{equation}
H^{\rm AKLT}=\sum_i P^{(J=2)}_{i,i+1},
\label{e:akltinfiniteprojection}
\end{equation}
where the operators $P^{(J=2)}_{i,i+1}=\fr{6}(\vec{S}_i \cdot \vec{S}_{i+1})^2 +\fr{2}(\vec{S}_i \cdot \vec{S}_{i+1}) + \fr{3} $, are projections onto the total spin-2 subspace of spin-1 particles $i$ and $i+1$,  where $\vec{S}$ is the spin-1 vector operator $(S_x,S_y,S_z)$, and the summation index $i$ goes over all integers.  This Hamiltonian is equivalent to that of Eq.~(\ref{e:akltinfinite}) up to an additive constant. As each $P^{(J=2)}_{i,i+1}$ operator is positive, a state that is a zero eigenstate of each $P^{(J=2)}_{i,i+1}$ will also be a ground state of $H^{\rm AKLT}$. The Hamiltonian $H^{\rm AKLT}$ is frustration free in the sense that there exists a state, the AKLT state, that minimizes the energy of each $P^{(J=2)}_{i,i+1}$ term separately.  

If we consider a finite $N$-particle chain with $i$ running from $1$ to $N$ we find that the ground state is four-fold degenerate. In this case, a unique state can be specified by appending spin-1/2 particles to the ends and adding the condition that the total spin of the end spin-1/2 particle and its neighboring spin-1 particle is 1/2.  Specifically, a Hamiltonian with the $N$-particle version of the AKLT state with attached spin-1/2 particles (which will be referred to simply as the AKLT state in the rest of this paper) as its ground state can be constructed analogously to above as a positive sum of projections. In terms of spin operators this Hamiltonian takes the form
\begin{equation}
H^{\rm AKLT}_N =\vec{s}_0\cdot \vec{S}_1 + \vec{S}_N\cdot \vec{s}_{N+1}+ \sum_{i=1}^N \vec{S}_i \cdot \vec{S}_{i+1} + \fr{3} (\vec{S}_i \cdot \vec{S}_{i+1})^2\,,
\label{e:akltfinite}
\end{equation}
where $\vec{s}$ is the spin-1/2 vector operator. The first two terms are projections onto total spin-3/2, and each summand is a projection onto total spin-2 (up to irrelevant additive constants and positive multiplicative factors). The above Hamiltonian is rotationally invariant and consists only of nearest-neighbour, two-body interactions~\cite{affleck1987}. It was proved in~\cite{affleck1988} that the Hamiltonian has a non-zero energy gap between the ground state and the first excited state. 

\begin{figure}
    \centering
    \includegraphics[width=0.45\textwidth]{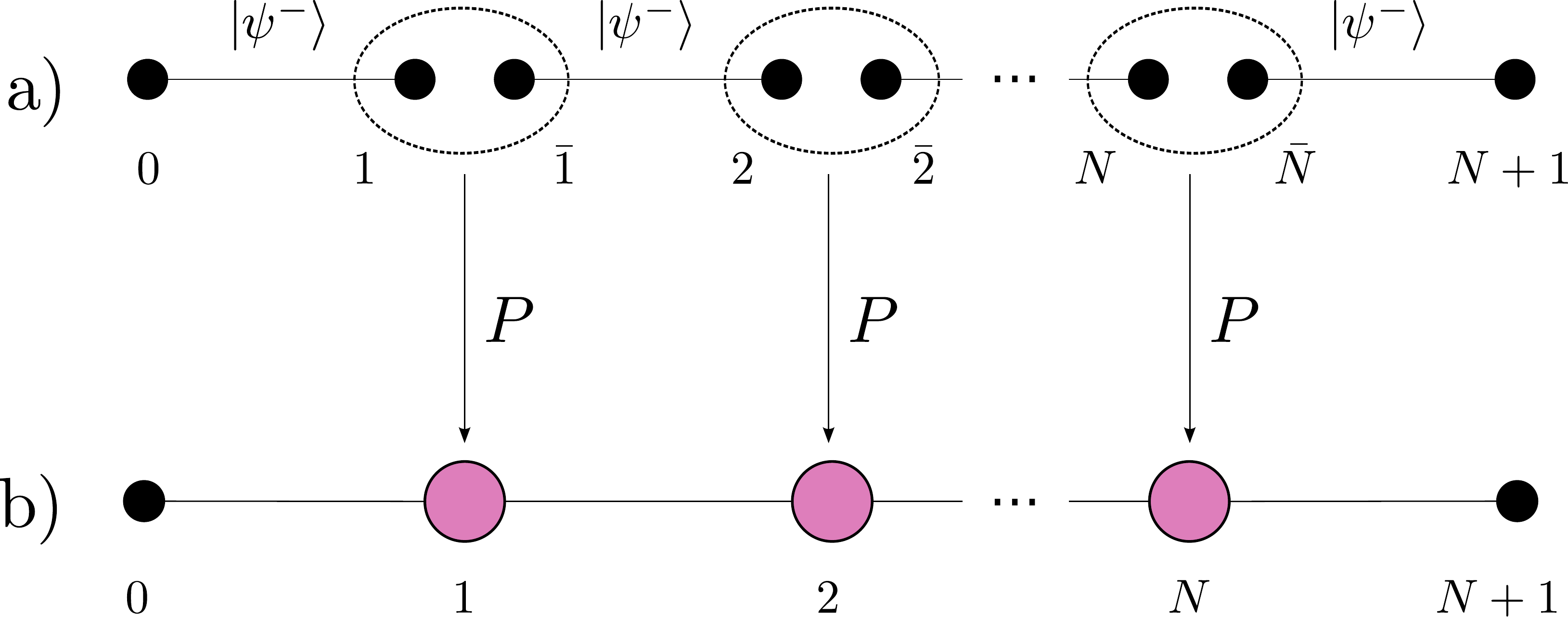}
    \caption{(Color online) Construction of the AKLT state from a line of singlets. Ends of singlets are `projected' onto spin-1 creating an entangled spin-1 chain.}
    \label{f:akltprojection}
\end{figure}

An explicit construction of the AKLT state is provided by its description as a valence bond solid. In this description, two ``virtual'' spin-1/2 particles are assigned to each spin-1 particle.  One is prepared in a spin singlet state with the neighbor to the left, and the other in a spin singlet with the neighbour to the right.  The pair of virtual pairs at each site are coupled to total spin 1.  Specifically, consider a line of spin-1/2 singlets $\sing=\rt{2}(\ket{01}-\ket{10})$, where $\ket{0}$ and $\ket{1}$ are spin-up and spin-down states, respectively, for the virtual spin-1/2 particles. Neighbouring singlets end on the sites where the physical spin-1 particles will be located, as in Fig.~\ref{f:akltprojection}a.

The pairs of virtual spin-1/2 particles at each site are projected onto the combined spin-1 subspace (the triplet) to create an entangled spin-1 chain, which is the ground state of Eq.~\eqref{e:akltfinite}, as in Fig. \ref{f:akltprojection}b. This construction using singlets (which have total spin 0) ensures that the total spin of any neighbouring spin-1 particles is not 2.

The AKLT state can then be expressed as
\begin{equation}
\ket{V}=(\otimes^N_{k=1} P_{k\bar{k}}) \ket{\psi^-}_{01}\ket{\psi^-}_{\bar{1}{2}}\cdots\ket{\psi^-}_{\bar{N}N+1}\,,
\label{e:akltstate}
\end{equation}
where $P_{k\bar{k}}$ is given by
\begin{equation}
P_{k\bar{k}}=\ketbra{M_1}{00}+\ketbra{M_0}{\psi^+} +\ketbra{M_{{-}1}}{11}\,,
\label{e:projection}
\end{equation}
and where $\ket{M_1},\ket{M_{{-}1}},\ket{M_0}$ are spin-1 eigenstates of $S_z$ (component of spin in the $z$ direction) and $\ket{\psi^+}=\rt{2}(\ket{01}+\ket{10})$. So the isometries $P_{k\bar{k}}$ project onto the total spin-1 subspace of a system of two spin-1/2 particles.

From this valence bond solid description, we may derive the matrix product state description of the AKLT state~\cite{verstraete2004b}
 \begin{equation}
\ket{V}=\sum_{\beta_1,\ldots,\beta_N} \ket{\beta_1}\cdots\ket{\beta_N} \eye \otimes A[\beta_N]\cdots A[\beta_1] \sing_{0,{N+1}}\,,
\label{e:akltmatrix}
\end{equation}
where $\eye$ is the $2\times2$ identity operator, $\ket{\beta_i}$ form any basis of the $i$-th spin-1 system $\mathcal{H}_i$ and the map $A: \mathcal{H}_i \rightarrow \mathfrak{sl_2(\mathbb{C})} $ (the space of traceless $2\times2$ matrices) is the bijective linear map satisfying
\begin{equation}
P_{i\bar{i}}(A[\beta]\da\otimes\eye)\sing_{i\bar{i}}=\ket{\beta_i}\,.
\label{e:matrix2spin1}
\end{equation}
For example, in the basis $\{ \ket{M_{{-}1}}, \ket{M_0}, \ket{M_1} \}$ we obtain the operators
\begin{align}
A[M_1]&=-\sqrt{2}\ketbra{1}{0}=-\sqrt{2}\sigma_-,\\
A[M_0]&=\ketbra{0}{0}-\ketbra{1}{1}=\sigma_z,\\
A[M_{{-1}}]&=\sqrt{2}\ketbra{0}{1}=\sqrt{2}\sigma_+.
\end{align}
Note that the singlet has the property that $\eye\otimes A\sing = \widetilde{A} \otimes \eye \sing$ where $A$ can be any $2 \times 2$ matrix and $\widetilde{A}=\sigma_y A^T \sigma_y $ where $\sigma_y$ is the Pauli $Y$ matrix.  This allows us to shift the $A$ operators from particle $N+1$ to particle $0$ depending on which is more convenient. 

\subsection{The AKLT state as a quantum computational wire}

We now demonstrate how the AKLT state can be used as a quantum computational wire, first shown in~\cite{gross2007} (where a minor variation on the AKLT state was used). The way in which information is transmitted along an AKLT state is analogous to teleporting a qubit multiple times. If the right measurements are performed on the AKLT state's spin-1 particles, a teleporting measurement can be realized on its underlying, virtual spin-1/2 particles. In Eq.~\eqref{e:akltmatrix} we have written the states of the spin-1 particles (labeled 1 to $N$) to the left of the spin-1/2 particles (labeled $0$ and $N+1$). The state is in an entangled superposition with the operators $\eye \otimes A[\beta_N]\cdots A[\beta_1]$ acting on particles $0$ and $N+1$ in each term. Performing measurements on every spin-1 particle in the basis $\beta$ will place the unmeasured spin-1/2 particles in a state of the form
\begin{equation}
\eye \otimes A[\beta_N]\cdots A[\beta_1] \sing_{0,{N+1}}\,,
\label{e:result}
\end{equation}
where the $\beta_i$ now label measurement outcomes. If each $A[\beta_i]$ is unitary, the spin-1/2 particles will be maximally entangled. It turns out that $A[\beta]$ will be unitary if and only if $\ket{\beta}$ is a zero eigenstate of spin along some physical axis. For instance, writing the AKLT state in Eq.~\eqref{e:akltmatrix} in the basis $\{\rt{2}(\ket{M_{{-}1}} + \ket{M_1}), \ket{M_0}, \rt{2}(\ket{M_{{-}1}}-\ket{M_1}) \}$, which are zero eigenstates of $S_y, S_z, S_x$,  will yield a Pauli operator for each $A$, specifically
\begin{align}
A\Bigl[\tfrac{1}{\sqrt{2}}\bigl(\ket{M_{{-}1}} + \ket{M_1}\bigr)\Bigr]&=i\sigma_y,\\
A[M_0]&=\sigma_z,\\
A\Bigl[\tfrac{1}{\sqrt{2}}\bigl(\ket{M_{{-}1}}-\ket{M_1}\bigr)\Bigr]&=\sigma_x.
\end{align}
The fact that particles $0$ and $N+1$ can be placed in a maximally entangled state, which can subsequently be used for teleportation, illustrates how the AKLT state has the capacity to transmit a qubit along its length. An alternate interpretation of Eq.~\eqref{e:result} is that we may perform a measurement on particle $0$ before measuring the spin-1 particles, and then the matrices $A[\beta]$ can be thought of as `acting on' particle $N+1$. We will elaborate on this idea for a linear optical implementation in the following sections.


%
%
%
%


\section{Optical implementation}
\label{s:optical}

We now show how to create an optical AKLT state with linear optical methods using entangled photon pairs, and subsequently use it as a quantum computational wire. Our proposed implementation encodes an AKLT state on an entangled chain of polarization biphotons, which serve as the spin-1 particles of the AKLT chain.  Biphotons are pairs of frequency degenerate photons occupying the same spatio-temporal mode with a polarization degree of freedom. Each biphoton is a three level system, a qutrit, spanned by the three states 
\begin{align}
\ket{HH} &:= \tfrac{1}{\sqrt{2}}\crusq{H}\ket{vac}\,,   \\
\ket{HV} &:= \cru{H}\cru{V}\ket{vac}\,,  \\
\ket{VV} &:= \tfrac{1}{\sqrt{2}}\crusq{V}\ket{vac}\,,
\end{align}
Note that, in our notation, the state $\ket{HV}$ is defined as a \emph{symmetric} state of two photons in the same spatio-temporal mode. In this paper we will regard spin-1/2 states as the horizontal and vertical polarization states of a single photon
\begin{equation}
\ket{0}=\cru{H}\ket{vac}\,, \qquad
\ket{1}=\cru{V}\ket{vac}\,,
\end{equation}
and the spin-1 states as the symmetric biphoton states
\begin{align}
\ket{M_1}&=\ket{HH}\,, \nonumber \\
\ket{M_0}&=\ket{HV}\,, \\
\ket{M_{{-}1}}&=\ket{VV}\,, \nonumber
\end{align}
where $\cru{H}, \cru{V}$ are the creation operators for horizontally and vertically polarized photons, respectively.

\subsection{Creating a photonic AKLT state}

\begin{figure}
    \centering
    \includegraphics[width=0.45\textwidth]{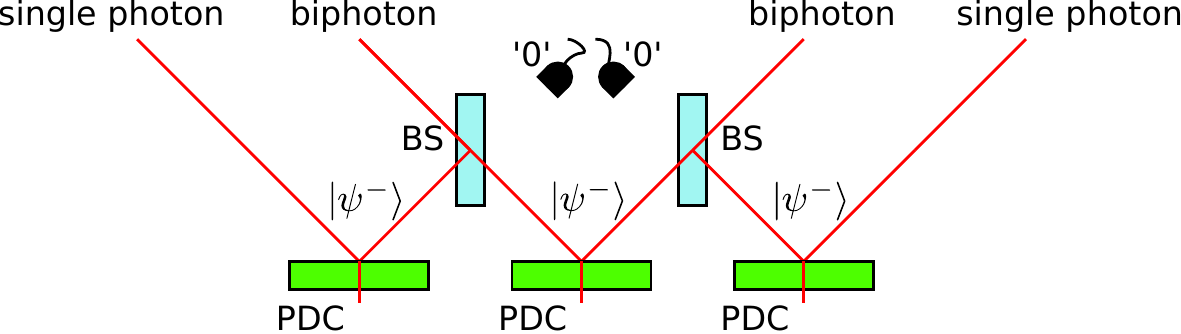}
    \caption{(Color online) An optical AKLT state with $N=2$. A post-selection of `0' counts at the photodetectors projects two photons, one from each neighbouring singlet, onto three level biphotons. Polarization states are created via parametric down conversion (PDC), and the two photons are interfered on 50:50 beam splitters.}
    \label{f:opticalaklt}
\end{figure}

We propose creating a photonic AKLT state following the PEPS construction described in the previous section. The method is illustrated in Fig.~\ref{f:opticalaklt}. In this construction, the AKLT state is built from an underlying line of singlets to which projections onto total spin 1 are applied at each site. The singlets can be physically realized by generating a line of type-II phase-matched parametric down-converted (PDC) photon pairs~\cite{kok2007} in polarization singlet configurations. The necessary projections $P_{k\bar{k}}$ can then be performed by passing two photons, one from each neighbouring singlet, through a 50:50 beam splitter so that they undergo Hong-Ou-Mandel interference~\cite{hong1987}. A well-known effect in quantum optics~\cite{mattle1996} is that when two photons of an incoming antisymmetric singlet state interfere on a 50:50 beam splitter, they will always emerge in separate arms. Conversely the outgoing photons of any symmetric input state will always emerge together in the same arm. Outputs for different beam splitter inputs are listed in Table~\ref{t:innsbruck}.  Thus by discarding outcomes in which photons emerge in separate arms, one projects out the singlet and ensures that both photons emerge as a biphoton. This is equivalent to applying the operator $P_{kk}$ to pairs of incoming polarized photons.

\begin{table}
\centering
\begin{tabular}{|c|c|}
\hline
Input state (Bell state)& State after beam splitter\\
\hline\hline
$(\cre{a}_V\cre{b}_V+\cre{a}_H\cre{b}_H)\ket{vac}$&$(\cre{b}_L\cre{b}_R - \cre{a}_L\cre{a}_R)\ket{vac}$\\
$(\cre{a}_V\cre{b}_V-\cre{a}_H\cre{b}_H)\ket{vac}$&$(\cre{b}_A\cre{b}_D - \cre{a}_A\cre{a}_D)\ket{vac}$\\
$(\cre{a}_V\cre{b}_H+\cre{a}_H\cre{b}_V)\ket{vac}$&$(\cre{b}_H\cre{b}_V - \cru{H}\cru{V})\ket{vac}$\\
$(\cre{a}_H\cre{b}_V-\cre{a}_V\cre{b}_H)\ket{vac}$&$(\cre{a}_H\cre{b}_V - \cre{a}_V\cre{b}_H)\ket{vac}$\\
\hline
\end{tabular}
\caption{Dependence of output on input photons. The creation operators $\cre{a}$ and $\cre{b}$ create photons in separate spatial modes and the subscripts $H, V, D, A, L, R$ denote horizontal, vertical, diagonal, antidiagonal, left and right circular polarizations respectively.}
\label{t:innsbruck}
\end{table}

The required postselection can be performed in more than one way. We could, for instance, postselect on one arm of the beam splitter. In this case we would place our detection apparatus on one arm and only regard the outcome as successful if two photons are detected on that arm. In theory this is equivalent to only accepting zero photons on the other arm. Such a set-up is illustrated in Fig.~\ref{f:opticalaklt}. In this case the probability of successfully adding a biphoton to the chain is $3/8$. We could double this probability by placing a detection apparatus on each arm of the beam splitter. In this case the only unsuccessful outcome is if one photon emerges on each arm, which is equivalent to detection of the singlet state. The probability of unsuccessful postselection at each projected site is 1/4, and thus the probability of successfully adding a biphoton to the chain is $3/4$. 

In terms of preparing a resource for MBQC, the latter process has several advantages compared with the `fusion' used to produce cluster states in linear optics~\cite{browne2005}.  First, the success rate is higher:  $3/4$ here compared with $1/2$ for fusion.  Second, a failure outcome corresponds to a projection onto a singlet state which, according to the rule of entanglement swapping 
\begin{equation}
_{23}\singb \left( \sing_{12}\sing_{34} \right)=-\fr{2}\sing_{14},
\end{equation}
removes two spin-1/2 particles (corresponding, in the successful case, to a single spin-1 particle) from the chain but entangles the next two particles, converting two singlet pairs into one.  Therefore, the failure outcome has no negative effect; the chain simply does not grow. 
After successful postselection, the resulting entangled line of biphotons will exactly encode the AKLT state. In the case where postselection is performed on both arms of the beam splitter, the average length of the chain (in terms of the number of spin-1 particles produced), starting with $N$ entangled photon pairs will be $3N/4-1$.  We now detail how such a state may be used as a wire for MBQC.  As we will show, the advantages in this approach for preparing an AKLT state compared with the cluster state are countered by more stringent requirements on the measurements needed to manipulate quantum information on the wire.

\subsection{Quantum computational wire operations}

Using an AKLT state to encode and manipulate a qubit relies on the ability to perform measurements on individual spin-1 particles. This capability is a major challenge in most atomic and condensed matter systems. However, in quantum optics, it is possible (and in fact straightforward) to perform high-fidelity projective measurements on single photons.  Bi-photon measurements are possible (although nontrivial) in our linear optical implementation, as we will discuss.

The logical qubit is encoded on the physical state of the measured AKLT state. This qubit evolves as single particle projective measurements are performed on the state. A helpful way of visualizing this is in terms of a correlation space~\cite{gross2007}. The correlation space is the space on which the matrices $A[\beta]$ act in the AKLT state's matrix product state description. A measurement on spin-1 particle $i$ will collapse the superposition in Eq.~\eqref{e:akltmatrix}, and fix the matrix $A[\beta_i]$ according to the measurement outcome $\beta_i$. Thus, successively measuring particles 1 through to $N$ will fix a sequence of $N$ matrices that act on the correlation space. By choosing different measurement bases, different operators can be applied to the correlation space. We will briefly outline how a qubit can be initialized, read out, or have an arbitrary qubit gate applied to it in this scheme.

\subsubsection{Qubit initialization}

There are two ways to initialize a correlation space qubit in an AKLT state of finite length, both of which are accessible with linear optical elements. One way is to perform a measurement on the spin-1/2 particle labeled 0. For each term in the summation of Eq.~\eqref{e:akltmatrix}, we have particles $0$ and $N+1$ existing in a singlet state with a product of matrices acting on particle $N+1$. Note that the singlet is antisymmetric, and so projecting the first qubit onto some state $\ket{s}$ will fix the state of the other particle as $\ket{s^\perp}$, the state orthogonal to $\ket{s}$. Hence, if we perform a measurement on particle 0 of the AKLT state and obtain an outcome of $\ket{s}$,  we will initialize particle $N+1$ in the state $\ket{s^\perp}$. This qubit, on which the matrices act, we regard as residing in the correlation space as discussed above. In our optical implementation, initializing the qubit in the state $\ket{0}$ or $X\ket{0}$ may be achieved by measuring the polarization of the end photon in the $\ket{H},\ket{V}$ basis. This measurement can easily be done by positioning a polarizing beam splitter in the path of the end photon, and counting the number of photons (1 or 0) appearing on each arm.
%
%
%

\begin{figure}
    \centering
    \includegraphics[width=0.25\textwidth]{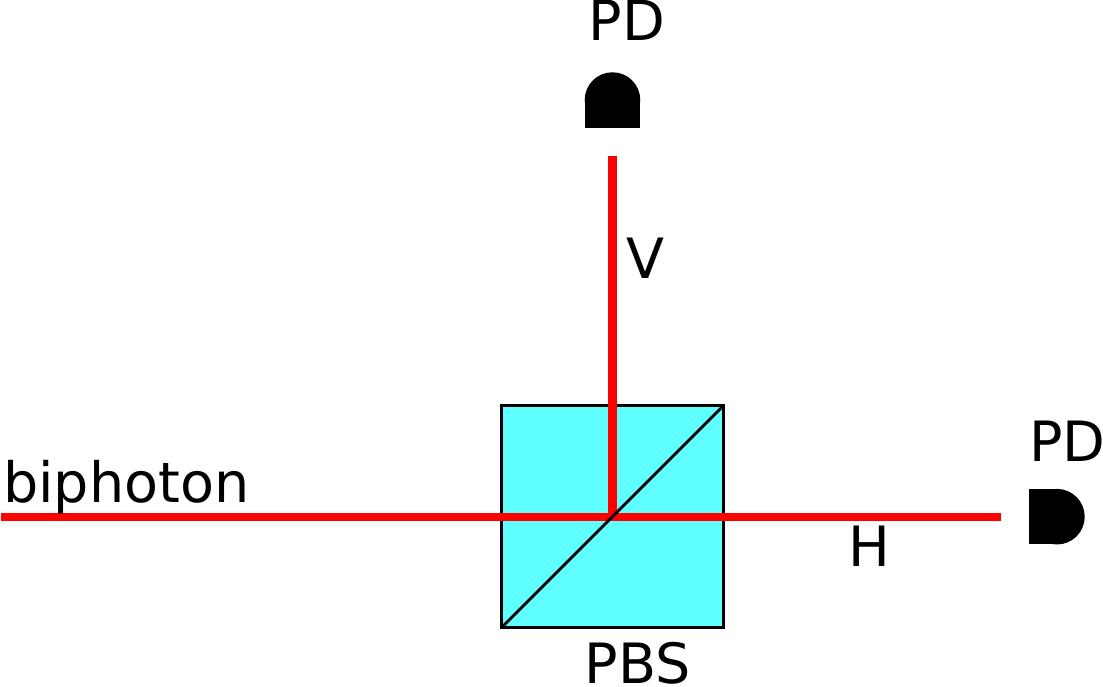}
    \caption{(Color online) The basic biphoton analyser, consisting of a polarizing beam splitter (PBS) and photodetectors (PD). Performing a measurement with the above device will initialize a qubit with probability 2/3. The three distinguishable outcomes are, two photons arrive at $V$, two photons arrive at $H$ and one photon arrives at each $V$ and $H$. Only the last of these will not initialize a qubit.}
    \label{f:measurementeasy}
\end{figure}

Alternatively, a qubit may be initialized in the correlation space by measuring any spin-1 particle in a disentangling basis where two of the three $A[\beta]$ operators are rank-1 (this is the maximum number of rank-1 operators possible in any given basis). Initialization will occur when an outcome corresponding to a rank-1 operator is obtained. For example, if a measurement of the $i$-th spin-1 particle is performed in the basis $\{\ket{HH}, \ket{HV}, \ket{VV}\}$, as illustrated in Fig.~\ref{f:measurementeasy}, then an outcome of $\ket{HH}$ (with corresponding operator $A[1]=-\sqrt{2}\ketbra{1}{0}=-\sqrt{2}\sigma_-$) will disentangle two halves of the AKLT state, transforming it to
\begin{multline}
\sum_{\beta_1,\ldots,\beta_{i-1}} \ket{\beta_1}\cdots\ket{\beta_{i-1}} \widetilde{A}[\beta_1]\cdots \widetilde{A}[\beta_{i-1}] \ket{1}_0 \otimes \ket{HH}_i \\
\otimes \sum_{\beta_{i+1},\ldots,\beta_N} \ket{\beta_{i+1}}\cdots\ket{\beta_N} A[\beta_N]\cdots  A[\beta_{i+1}] \ket{1}_{N+1} \,.
\end{multline}
We have written the two spin-1/2 particles to the right of the operators that act on them. Particles $0$ through to $i-1$ are completely disentangled from particles $i+1$ through to $N+1$. We regard this operation as initializing two qubits in the state $\ket{1}$ in two separate halves of the chain. An analogous result will hold if an outcome of $\ket{VV}$ is obtained (except the qubit will be initialized in the state $X\ket{1}$). If an outcome of $\ket{HV}$ is obtained then we replace the matrix $A[\beta_i]$ with a $Z$ operator. While this $Z$ operator is harmless (it may be compensated for in subsequent operations), no qubit will initialized in this case. Hence qubit preparation with this method is non-deterministic. A repeat-until-success strategy on successive particles may still be used to prepare a qubit, with probability 2/3 for each attempt.

\subsubsection{Readout of logical qubits}

The procedure for logical qubit readout is analogous to initialization. If every spin-1 particle has been measured, and the correlation space evolution is complete, then the state of the correlation space qubit will be encoded on the last unmeasured spin-1/2 particle. Readout may then be performed on this particle by direct measurement. A readout of the correlation space qubit can also be achieved by performing measurements on unmeasured spin-1 particles. For example, if the correlation space is in the state $A[\beta_{i}]\cdots A[\beta_{1}]\ket{0}_0=\alpha\ket{0}+\beta\ket{1}$ (so particles $0$ through to $i$ have been measured) then readout in the computational basis may be performed by measuring the next spin-1 particle in the basis $\{\ket{HH},\ket{HV},\ket{VV}\}$. An outcome of $\ket{HV}$ does not correspond to a readout, but rather performs a logical $Z$ Pauli operator to the correlation space. Note that the modulus of the $\alpha$ and $\beta$ coefficients is unaffected by this operation.  On the other hand, the probabilities of obtaining outcome $\ket{HH}$ and $\ket{VV}$ conditional on $\ket{HV}$ not being detected can be shown to be $|\alpha|^2$ and $|\beta|^2$ respectively. Thus this type of readout faithfully preserves measurement statistics. As the $\ket{HV}$ outcome does not correspond to a successful readout, this measurement scheme is non-deterministic. A readout can be performed by repeatedly measuring successive particles until a successful outcome (corresponding to successful readout) is obtained.


\subsubsection{Teleportation and unitary operations}


We now illustrate how to choose a biphoton basis which will apply a desired unitary operation to the correlation space using the explicit examples of the `identity' operator as well as the $Z$ and $X$ rotations. Keep in mind that we are just finding a bases with such a property, and it should not be assumed that performing a measurement in these bases is actually possible using only linear optics (it is not). We will address such problems in the next section. The basic idea of using the AKLT state to perform arbitrary unitaries was first illustrated in~\cite{gross2007}, however the approach we present follows that of~\cite{brennen2008}. 

A set of matrices that is typically used to characterize the AKLT state (in terms of its matrix product state description) are the Pauli matrices $X,Y,Z$. The physical basis corresponding to these matrices is the Bell basis excluding the singlet as can be found by substituting into Eq.~\eqref{e:matrix2spin1}, and corresponds to the bi-photon basis $\{ \ket{HV}, \ket{DA}, \ket{RL}\}$.  We refer to such a basis as a \emph{spin-0 basis}. We have explicitly written the elements of this basis with the corresponding Pauli operators in table~\ref{t:outcomeoperator}. When a measurement is performed in this basis, one of the three Pauli operators will be applied to the correlation space dependent on the measurement outcome. We call this applying the logical identity with Pauli `biproducts'. Biproduct operators are harmless in the sense that they generate a finite group (the Pauli group), and can be accommodated using the standard techniques of a transforming Pauli frame used in measurement-based quantum computation~\cite{raussendorf2001}. 

The bases required for performing $Z$ and $X$ rotations are similar to the above spin-0 basis.  First consider the $Z$ rotation 
\begin{equation}
Z(\theta)=e^{-\frac{i\theta}{2}}\ketbra{0}{0}+e^{\frac{i\theta}{2}}\ketbra{1}{1}.
\end{equation}
The $A[\beta]$ matrices that act on the correlation space are traceless for any measurement basis. Thus we should decompose $Z(\theta)$ (which is not, in general, traceless) into a product of a traceless operator with some biproduct ``error" operator. One example of such a decomposition is 
\begin{equation}
Z(\theta)= X \left( e^{\frac{i\theta}{2}}\ketbra{0}{1} + e^{-\frac{i\theta}{2}}\ketbra{1}{0} \right).
\end{equation}
The Pauli $X$ operator may be regarded as a biproduct operator. The measurement outcome $\ket{\beta_1}$ corresponding to $A[\beta_1]:=e^{\frac{i\theta}{2}}\ketbra{0}{1} + e^{-\frac{i\theta}{2}}\ketbra{1}{0}=XZ(\theta)$ can be found by substituting into Eq.~\eqref{e:matrix2spin1} to be 
\begin{equation}
\ket{\beta_1}=\rt{2}(\ket{HH} - e^{-i\theta}\ket{VV}).
\end{equation}
An alternative way of writing this state in terms of creation and annihilation operators is as 
\begin{equation}
(\crusq{H} - e^{-i\theta}\crusq{V})\ket{vac}=(\cru{H}+e^{-\frac{i\theta}{2}}\cru{V})(\cru{H}-e^{-\frac{i\theta}{2}}\cru{V})\ket{vac},
\end{equation}
where the right hand side clearly illustrates the fact that the biphoton contains two orthogonal photons.

We find our other basis elements by doing a second decomposition of $Z(\theta)$ into traceless operators
\begin{align}
Z(\theta) =XZ \left( e^{\frac{i\theta}{2}}\ketbra{0}{1} - e^{-\frac{i\theta}{2}}\ketbra{1}{0} \right).
\end{align}
The biphoton $\ket{\beta_2}$ corresponding to the operator $A[\beta_2]:=e^{\frac{i\theta}{2}}\ketbra{0}{1} - e^{-\frac{i\theta}{2}}\ketbra{1}{0}=ZXZ(\theta)$ is 
\begin{equation}
\ket{\beta_2}=\rt{2}(\ket{HH}+e^{-i\theta}\ket{VV}).
\end{equation}
Thus, a basis for performing $Z(\theta)$ rotations with Pauli biproducts can be chosen, where the third basis element is specified by the first two. This basis is listed in table \ref{t:outcomeoperator}.  The first two measurement outcomes apply the $Z(\theta)$ rotation with $X$ or $ZX$ biproducts to the correlation space.  The last outcome does not apply a rotation at all, but only a harmless $Z$ biproduct.  As described in~\cite{gross2007}, obtaining this ``failure'' outcome (which occurs 1/3 of the time) is heralded, and the rotation gate can be attempted again on the next spin-1 particle.  The rotation can then ultimately be achieved with arbitrarily high probability, given enough attempts.

\begin{table}
\centering
\begin{tabular}{|c|c|c|}
\hline
&Measurement outcome&Correlation operator\\
\hline \hline
\multirow{3}{*}{Identity} & $\ket{HV}$ & $Z$\\
&$\ket{DA}$ & $X$\\
&$\ket{RL}$ & $XZ$\\
\hline
\multirow{3}{*}{Z-rotation} & $\rt{2}(\ket{HH} - e^{-i\theta}\ket{VV})$ & $XZ(\theta)$\\
&$\rt{2}(\ket{HH} + e^{-i\theta}\ket{VV})$ & $ZXZ(\theta)$\\
&$\ket{HV}$ & $ Z$\\
\hline
\multirow{3}{*}{X-rotation} & $\rt{2}(\ket{DD} - e^{-i\theta}\ket{AA})$ & $ZX(\theta)$\\
&$\rt{2}(\ket{DD} + e^{-i\theta}\ket{AA})$ & $XZX(\theta)$\\
&$\ket{DA}$ & $X$\\
\hline
\end{tabular}
\caption{Measurement outcomes and their corresponding correlation space operators. The first three outcomes correspond to measurement in the `standard' basis, where every correlation space operator is a Pauli operator. The next three outcomes form the basis used for a $Z$ rotation. The last three form the basis used for an $X$ rotation.}
\label{t:outcomeoperator}
\end{table}

To perform an $X$ rotation, the method is similar to that of the $Z$ rotation and is obtained by exchanging the logical states $\ket{0}$ and $\ket{1}$ with the $X$ eigenstates $\rt{2}(\ket{0}\pm\ket{1})$ in all of the previous derivations. In terms of photons this simply corresponds to replacing the $H,V$ labels with the diagonal, antidiagonal labels $D,A$. Note also that the error operators $X$ and $Z$ which appear in the $Z$ rotations are swapped. The basis for performing $X$ rotations is listed in table~\ref{t:outcomeoperator}.

We now consider the form of the measurements used for the identity operator, $X$ and $Z$ rotations (and in fact for any unitary operator).  These measurements must be performed in a biphoton basis for which each element is a zero eigenstate of spin along some axis; thus the notation ``spin-0 basis''. In any spin-0 basis, the matrices $\{A[\beta_1],A[\beta_2],A[\beta_3]\}$ are equivalent to the three Pauli operators up to conjugation by a unitary matrix.  In order to restrict the biproduct operators to a finite group, we restrict our spin-0 bases to those presented in table~\ref{t:outcomeoperator}, and using these an arbitrary single qubit unitary may be realized via an appropriate sequence of measurements. To see this, first note that any single qubit unitary can be expressed as a product of three rotations $Z(\theta_3)X(\theta_2)Z(\theta_1)$. The $X$ and $Z$ rotations can be separately realized up to Pauli biproducts by performing measurements in the bases listed in table~\ref{t:outcomeoperator}. If the outcome only induces a biproduct and not a rotation, which occurs with probability 1/3 when the outcome in the last row is obtained, the same measurement can be repeated until a desired rotation outcome is obtained.  All of the biproducts, that depend on the measurement outcomes, can be brought out the front of the rotations using the relations $XZ(\theta)=Z(-\theta)X$ and $ZX(\theta)=X(-\theta)Z$. Feed-forward of measurement outcomes is required for this where, based on the knowledge of previous measurement outcomes, the measurement angle $\theta$ of subsequent measurements is changed to either $\pm\theta_i$ depending on what Pauli operator must be brought through. In this procedure, the length of the computation is inherently random, however any single qubit unitary can be realized with sufficiently many measurements.

\subsubsection{Spin-0 basis measurements with linear optics}

\begin{figure}
    \centering
    \includegraphics[width=0.25\textwidth]{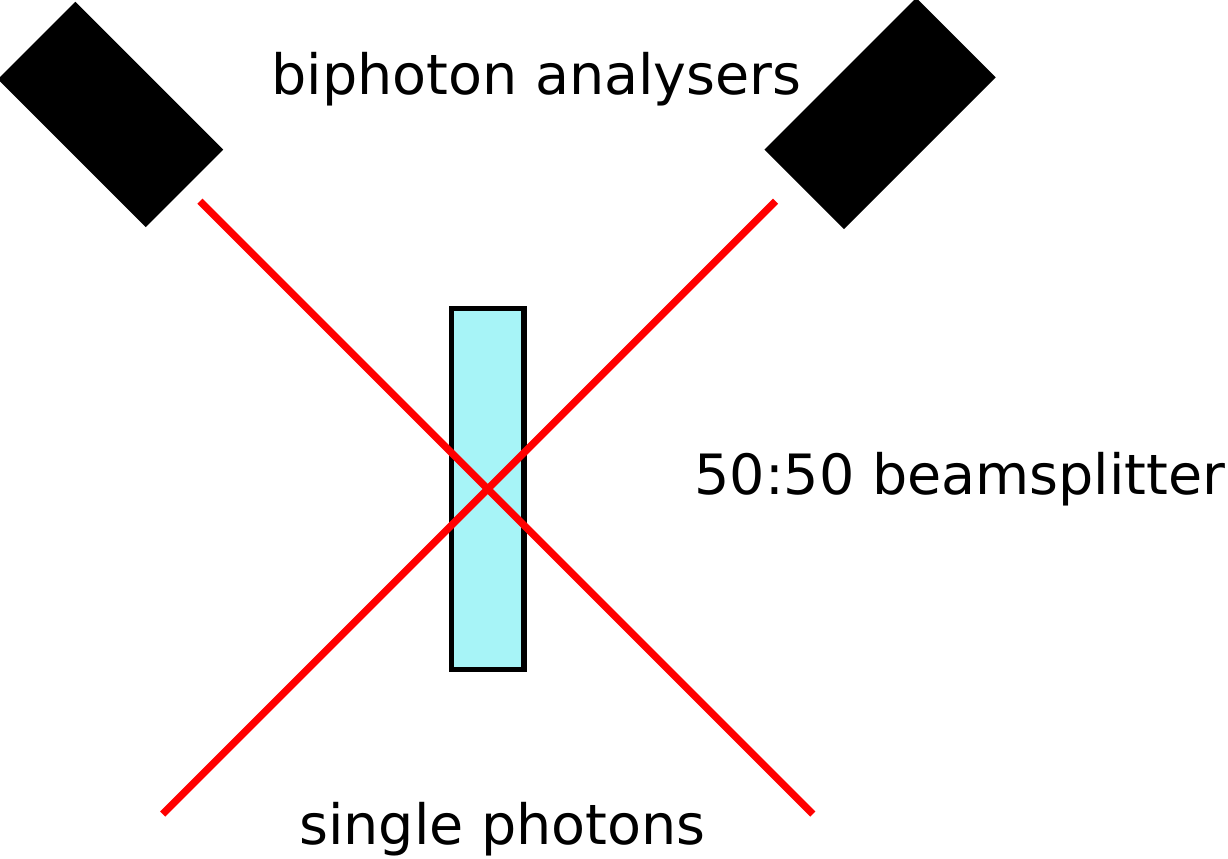}
    \caption{(Color online) The standard Innsbruck scheme for Bell measurements. If the biphotons could be detected with in a zero polarization degree basis, then deterministic Bell measurements would be possible.}
    \label{f:innsbruck}
\end{figure}

The biphoton measurements described in the previous section are challenging; as we know show, it is not possible to perform complete measurements in a spin-0 basis for biphotons using only linear optical methods. A spin-0 basis corresponds, in our proposed implementation, to a basis where each biphoton has zero polarization degree, i.e., each biphoton contains two photons with orthogonal polarizations. Each of the bases listed in table~\ref{t:outcomeoperator} have this property. The fact that such a measurement cannot be performed in linear optics is closely related to the problem of performing Bell measurements in linear optics~\cite{lutkenhaus1999}. In fact, we can place upper bounds on biphoton detection using the bounds for Bell measurements. Consider the standard Innsbruck detection scheme~\cite{weinfurter1994}, illustrated in Fig.~\ref{f:innsbruck}.
We have already listed the action of the 50:50 beam splitter on incoming Bell states, encoded on separate beam splitter arms, in Table~\ref{t:innsbruck}.  When either of the three symmetric Bell states are input, an output consisting of a zero polarization degree biphoton superposed in both arms is obtained. When a singlet is input, the photons emerge in separate beam splitter arms.

Let us assume that a linear optical measuring device is placed on each output arm of the beam splitter. An incoming singlet is heralded by a single photon count on each arm of the beam splitter. This will occur with probability 1/4 for maximally mixed input. If the measuring device that we placed on each output arm of the beam splitter could deterministically distinguish between the three zero polarization degree biphotons corresponding to the three symmetric Bell states, then we could deterministically perform a Bell measurement. However, the no-go theorem for Bell measurements~\cite{lutkenhaus1999} says that it is impossible to perform a measurement that distinguishes Bell states with certainty using linear optics alone, even allowing the use of feed-forward and auxiliary photons (note, however, success probabilities may be improved arbitrarily at the cost of more auxiliary input photons e.g., in the KLM protocol~\cite{kok2007}). Hence we cannot distinguish between three orthogonal zero polarization biphotons using feed-forward and auxilliary photons because if we could, we could also distinguish Bell states deterministically.

Also note a simpler version of the no-go theorem where no feed-forward or auxiliary photons are used. In this case the probability of obtaining a Bell state outcome given maximally mixed input cannot exceed 1/2. If we have a biphoton analyser that projects onto a zero polarization degree biphoton with probability $1/3$ then we can saturate this probability by placing it on one of the outgoing arms of the beam splitter in Fig.~\ref{f:innsbruck}. We will then have a probability of $1/4$ of projecting onto a singlet and a probability of $1/3\times 3/4=1/4$ of detecting a biphoton corresponding to a Bell state, giving the total probability of projecting onto a Bell state of $1/2$. 
The basic polarization analyser illustrated in Fig. \ref{f:measurementeasy} saturates the probability of zero polarization degree biphoton detection. Coincidence detection at the two detectors projects onto a zero polarization degree biphoton, and this happens with a probability of 1/3 for maximally mixed input. The illustrated set-up projects onto the $\ket{HV}$ biphoton, however with the addition of waveplates this biphoton can be changed arbitrarily to any biphoton of zero polarization degree. Hence we can project onto any of the biphotons in table \ref{t:outcomeoperator} and thus any of the correlation space operators $A[\beta]$ in the second column of table \ref{t:outcomeoperator} can be realized. 
However, using this measurement to apply unitaries to the correlation space is nondeterministic. An undesirable measurement outcome, corresponding to a double count at either of the photo detectors, will apply a rank-1 operator to the correlation space, collapsing the state of the correlation space qubit. The probability of obtaining a zero polarization degree photon is $1/3$ for each measurement, thus the probability of successful state transfer along the AKLT state diminishes by a factor of $1/3$ for each measured spin-1 particle.

Despite the non-existence of a simple, deterministic linear optical scheme for performing these measurements, one could investigate the use of techniques from linear-optical quantum computing~\cite{kok2007} to use ancilla photons and single-photon measurement to induce the nonlinearity needed for such measurements.  The success of such schemes in performing Bell state analysis with linear optics~\cite{langford2005} suggest that similar schemes may exist for spin-0 basis measurements of biphotons.  Finally, we note that MBQC schemes with single photons in general have very stringent requirements on the measurements; current photodetectors are not yet able to meet the efficiency thresholds for fault-tolerant MBQC with optics including cluster-state schemes.  Potentially, in the development of novel detection methods with ultrahigh efficiency (for example, based on the high-efficiency transfer of optical quantum information into atomic or solid state devices required for quantum repeaters), the nonlinear measurements required for MBQC using an optical AKLT state may indeed be possible.

\section{Conclusion}
\label{s:conclusion}

We have shown how an AKLT state may be realized in linear optics, and how elementary MBQC operations, including state preparation, measurement, and unitary logic gates, can be performed using this AKLT state as a resource.  The method we use to construct the AKLT state is inspired by its VBS construction: starting with a line of photon pairs in polarization singlet states, we apply projections onto total spin-1 by interfering pairs of photons, one from each neighbouring singlet, on a 50:50 beam splitter and then postselect. The AKLT state will then be encoded on an entangled line of biphotons. The success probability of adding a single spin-1 particle to the photonic AKLT state can be $3/4$. We also showed how wire operations may be applied via measurement using basic polarization analyser made of photodetectors and polarizing beam splitters, including initialization, readout and the application of arbitrary single qubit unitary operators. 

Our proposal demonstrates how MBQC may be performed on a state from condensed matter physics that is different from the cluster state, and which leads to different requirements.  Compared with a cluster state, the optical AKLT state is significantly simpler to create; however, its capacity for quantum computation in linear optics is more restrictive due to limitations of biphoton detection in linear optics.  As biphotons cannot be measured in an arbitrary basis, one cannot deterministically perform the measurements required to implement unitary gates.  An arbitrary single qubit unitary can only be applied nondeterministically with linear optics. These issues highlight the restrictive nature of biphoton measurement, and motivate the development of techniques for biphoton measurement within a linear optical setting. 

\begin{acknowledgments}
We thank Andrew Doherty, Rainer Kaltenbaek, Anthony Laing, Jonathan Lavoie, Jeremy O'Brien, Geoff Pryde, Kevin Resch, Terry Rudolph, Howard Wiseman, and Bei Zeng for discussions.  SDB acknowledges the support of the Australian Research Council.
\end{acknowledgments}


\end{document}